\documentclass[aip]{revtex4}
\usepackage{dcolumn}
\usepackage{amsmath}
\usepackage{amssymb}

\usepackage[dvipdfmx]{graphicx}
\usepackage{epstopdf}
\usepackage{subfig}

\newcommand{\mat}{\mathbf}
\renewcommand{\vec}{\mathbf}

\begin{document}

\title{Finding communities in sparse networks}
\author{Abhinav Singh}

\email{abhinav.singh@manchester.ac.uk}
\author{Mark Humphries}
\affiliation{University of Manchester}

\begin{abstract}
Spectral algorithms based on matrix representations of networks are often used to detect communities but classic spectral methods based on the adjacency matrix and its variants fail to detect communities in sparse networks. New spectral methods based on non-backtracking random walks have recently been introduced that successfully detect communities in many sparse networks. However, the spectrum of non-backtracking random walks ignores hanging trees in networks that can contain information about the community structure of networks. We introduce the reluctant backtracking operators that explicitly account for hanging trees as they admit a small probability of returning to the immediately previous node unlike the non-backtracking operators that forbid an immediate return. We show that the reluctant backtracking operators can detect communities in certain sparse networks where the non-backtracking operators cannot while performing comparably on benchmark  stochastic block model networks and real world networks. We also show that the spectrum of the reluctant backtracking operator approximately optimises the standard modularity function similar to the flow matrix. Interestingly, for this family of non- and reluctant-backtracking operators the main determinant of performance on real-world networks is whether or not they are normalised to conserve probability at each node.
\end{abstract}
\maketitle

Many networks have a modular structure. Social networks contain communities of friends \cite{zachary1977information,Girvan2002,Newman2006b}, collaborators \cite{Girvan2002}, and dolphins \cite{Lusseau2004}; brain networks contain groups of correlated neurons \cite{Harris2003,Humphries2011}, circuits of connected groups \cite{Binzegger2004,Perin2011}, and regions of connected circuits \cite{Meunier2009}. Similarly modular networks occur across biological domains from protein interaction networks to food webs \cite{Fortunato2010}. This range of applications has driven the dramatic development of ``community detection" methods for solving the core problem of finding modules within an arbitrary network  \cite{Fortunato2010}. Especially popular are spectral methods based on the eigenvalues and eigenvectors of some matrix representation of the network. These combine speed of execution with considerable information about the network beyond the modular structure \cite{newman2006finding}, including the relative roles of each node \cite{newman2006finding} and characterisation of the network's dynamical properties \cite{Rajan2006,Zhou2009}.

Spectral methods can fail for a range of real networks. These methods rely on the eigenvalues falling into two classes, the vast majority -- the ``bulk" -- following a well-defined distribution, and the outliers from that distribution giving information about the modular structure. Topological features of a network unrelated to its modules, such as network hub nodes with high degree, can distort this distinction by introducing eigenvalues outside the bulk that mix with those containing information about modules\cite{Farkas2001,Goh2001,Nadakuditi13}. Sparse networks often contain such network hubs and the outlying uninformative eigenvalues cause the breakdown of spectral methods \cite{Zhang2012}. Unfortunately many real-world networks are sparse (see Table II in \cite{Newman2003b} and Table 1 in \cite{Humphries2008}).

Krzakala et al. \cite{krzakala2013spectral} proposed a new ``non-backtracking" matrix representation of a network that solves this problem : their matrix represents a random walker on the network who cannot immediately return to a node it has just left. The eigenspectrum of the adjacency and related matrices is closely related to properties of a random walker traversing a network. In particular, the eigenspectrum  depends on the frequency with which the walker passes through any given node. As the non-backtracking matrix forbids the random walker to return to its immediately previous node, network hubs are not visited disproportionately by this random walker and  the spectrum of this random walker is not distorted by the presence of hubs in the network. Following this, Newman introduced the closely-related ``flow" matrix \cite{newman2013spectral} that conserved the probability for the random walk. Spectral methods applied to these matrices successfully recover modules in sparse networks, down to the theoretical limit for their detection in classes of model networks \cite{krzakala2013spectral}.

However, as noted by Newman \cite{newman2013spectral}, these represent an incomplete solution as networks containing trees cannot be handled elegantly. Because the random walker could not escape from such a tree once entered, trees are ignored despite being candidates for separate modules. In this paper we introduce the ``reluctant backtracker" approach, which combines the advantages of these new matrix representations by retaining the power of spectral methods for sparse networks with the ability to detect and correctly handle networks with trees. We show that this comes with no penalty for detection performance compared to non-backtracking and flow matrices. Rather, we show that the main difference in performance depends on whether or not such matrix representations are normalised to conserve probability. This finding hints at some deeper difference in network structure than modularity alone.

\subsection*{Non-backtracking and flow matrices}
We first outline the non-backtracking \cite{krzakala2013spectral} and flow matrix \cite{newman2013spectral} approaches. Both these approaches and ours start from the same representation of the network. Assume an unweighted undirected connected network with $n$ vertices and $m$ edges without self loops. We convert the undirected network into a directed network with $2m$ edges by replacing the undirected edge with directed edges in both directions; $j \to i$ showing the direction of the edge. The binary non-backtracking matrix $\vec{B}$ has $2m \times 2m$ elements, each element corresponding to a pair of directed edges in the network. Its elements are given by

\begin{equation}
B_{j\to i,l\to k} = \delta_{il}(1-\delta_{jk})
\end{equation}

The non-backtracking matrix is a sparse binary matrix. Elements are non-zero only if $B_{j\to i,l\to k}$  corresponds to a directed path from  $j$ to $k$ that passes through node $i$ with the restriction that nodes $j$ and $k$ must not be identical, i.e. no  backtracking. This matrix encapsulates the biased random walker that is prohibited from returning to its immediately previous node.

Newman modified the non-backtracking matrix by changing the values of its non-zero elements and called it the flow matrix $\vec{F}$. The matrix $\vec{F}$ is called the flow matrix based on an analogy with current flow in an electrical network. Its elements are given by
\begin{equation}
F_{j\to i,l\to k} = \delta_{il}(1-\delta_{jk}) \frac{1}{d_i -1},
\end{equation}

where $d_i$ is the degree of the node $i$. Consider the random walker that starts from node $j$ and is passing through node $i$. According to the flow matrix, the random walker is can reach any of the $d_i -1$ nodes except node $j$ with equal probability. The probability of reaching node $k$ from node $j$ passing through node $i$ is $\frac{1}{d_i -1}$. Therefore, probability is conserved at node $i$ just like current is conserved at each node in an electrical network; the amount of current entering a node must be equal to the current leaving a node. Krzakala et al.~\cite{krzakala2013spectral} and Newman~\cite{newman2013spectral} respectively showed that the 2nd leading eigenvector of the non-backtracking and flow matrices is very successful in correctly dividing sparse networks into modules.

\section*{Results}

\subsection*{Reluctant backtracking operators}

To solve the problem of detecting modules in the presence of trees, we introduce the idea of a reluctant backtracking random walker that admits a small probability of returning to a node immediately. The reluctance, but not impossibility, of immediately returning to a node mitigates network hub effects on the spectrum of the operators, while allowing the walker to explore and return from hanging trees unlike the non-backtracking operator or flow matrix.

Based on this idea of reluctance, we define two new reluctant backtracking operators $\mat{R}~\textrm{and}~\mat{P}$ whose matrix elements are
\begin{align}
\mat{R}:~  R_{j\to i,l\to k}= & \delta_{il}(1-\delta_{jk}) + \delta_{il}\delta_{jk}\frac{1}{d_j}\\
\mat{P}:~  P_{j\to i,l\to k}= & \left[\delta_{il}(1-\delta_{jk}) + \delta_{il}\delta_{jk}\frac{1}{d_j}\right]\frac{1}{d_i -1 + \frac{1}{d_j}}
\end{align}

where $R_{j\to i,l\to k}$ and $P_{j\to i,l\to k}$ represents the probability that the random walker shall move from node $j$ to node $k$ with nodes $i$ and $l$ as intermediate nodes. The probability of returning to a node for both operators $\mat{R}~\textrm{and}~\mat{P}$ is inversely proportional to the degree of the node, thus discouraging strongly a return to a high degree node.

The operator $\mat{R}$ is a reluctant version of the non-backtracking operator $\mat{B}$ as it allows the additional probability $\frac{1}{d_j}$ of returning immediately to the node $j$. The operator $\mat{P}$ is a normalised version of the operator $\mat{R}$ just like the flow operator $\mat{F}$ is a normalised version of the non-backtracking operator $\mat{B}$. The probability of  reaching a node is equal to the probability of leaving a node akin to the conservation of current at each node in an electrical network for the normalised operators $\mat{P}$ and $\mat{F}$.

The procedure for detecting the communities is identical for both operators. Given the adjacency matrix of a network, we first generate one of the matrices $\mat{R}~\textrm{or}~\mat{P}$. Following  Krzakala et al. \cite{krzakala2013spectral}, we calculate its second largest absolute real eigenvalue and the associated eigenvector. The eigenvector has $2m$  elements corresponding to each directed edge in the network. We group the elements of the eigenvector by the group index of the source node of each edge and sum them up to create a new vector that has $n$ elements corresponding to each node in the network. We  divide the network into two communities by grouping all nodes that have the same sign; the sign of each element represents the estimate of the reluctant backtracking operators of the node's community.

\subsection*{Communities composed of trees}
The indifference of  non-backtracking operators towards trees can impair their abilities to detect communities in networks. As an extreme case, consider the  network suggested by Newman~\cite{newman2013spectral}: a network composed of two binary trees connected at a single node. The non-backtracking operator $\mat{B}$ and the flow matrix $\mat{F}$ cannot detect communities in such a network, but the reluctant backtracking operators $\mat{R}$ and $\mat{P}$ do.

\begin{figure}[htbp] 
   \includegraphics[width=0.5\columnwidth]{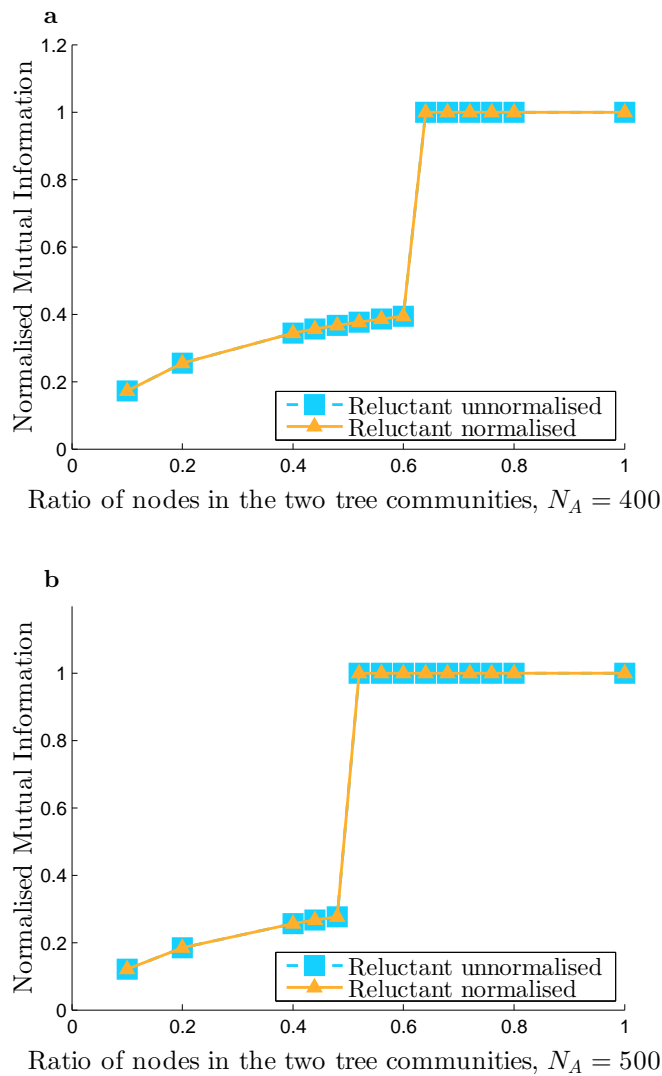}\\

   \caption{Two binary trees connected at one node. The x-axis shows the number of nodes in Community $B$ as a fraction of nodes in community $A$. The triangles and squares  show the performance of the two operators in detecting communities as measured by the normalised mutual information (NMI): $0 \leq NMI \leq 1 $, where $NMI =1$ means perfect community detection and $NMI =0$ means random allocation of nodes to communities (see Methods for more details). (a) 400 nodes in community $A$. Number of nodes in community $B$ varies from from 40 to 400. (b) 500 nodes in community $A$. Number of nodes in community B varies from 50 to 500.}

   \label{fig:example}
\end{figure}

We show this using a network composed of two communities $A$ and $B$ where each community is a tree and the two communities are connected by a \textit{single} node. The ratio of number of nodes in community $A$ and $B$ is denoted by $f$. Figure~\ref{fig:example} shows the performance of the reluctant backtracking operators in detecting the two communities in these simulated networks. The number of nodes in community $A$ is fixed at 400 and 500  in Figure~\ref{fig:example} panel (a) and panel (b) respectively, and the number of nodes in community $B$ varies. When the size of the two communities is comparable, the reluctant operators detect communities perfectly since the two communities are almost disconnected except one connection between the two communities and random walkers remain within the same community for substantial periods of time. There is a sharp transition in the ability of the reluctant backtracking operators around $f = 0.6$ in the network where community $A$ consists of 400 nodes. When one community becomes much smaller than the other, random walkers keep moving to the larger community from the small community in a short period of time and leads to the loss of performance. There is nothing universal  about the transition point fraction $f= 0.6$ as the transition point is a function of the number of nodes in the network and changes to $f=0.48$ when community $A$ has 500 nodes. The exact nature of transition in performance around the transition point $f$ is dependent on many factors such as the structure of the network, total number of nodes in the network and the relative sizes of different communities.

\subsection*{Stochastic block model with additional leaves}
Networks composed solely of trees are of course very artificial, but we also show that reluctant backtracking operators can detect communities in a more plausible network where the non-backtracking operators fail. Consider a more typical network, created by the classic stochastic block model. The addition or deletion of hanging trees to this network or any other will not affect the eigenspectrum of the non-backtracking operator $\mat{B}$. However, the presence/absence of hanging trees can significantly alter the structure of communities in such a network.

Stochastic block models provide an easy recipe for constructing networks with specified inter-community and intra-community edge probability. Consider a network of $n$ nodes with two communities. The probability of an edge between nodes $i$ and $j$ is given by

\begin{align}
\label{}
   P_{ab}= &\frac{c_{in}}{n} &\text{if a and b belong to same community}   \\
   =&\frac{c_{out}}{n} &\text{if a and b belong to different communities}
\end{align}

Let $c=\frac{c_{in} + c_{out}}{2}$ be the average degree of the network and $c_{minus} = \frac{c_{in} - c_{out}}{2}$ denote the degree of mixing between communities in the network. No mixing between the communities implies $c_{minus}=c$ and complete mixing between the 2 communities implies $c_{minus}=0$.

We demonstrate the effect of hanging trees by selectively adding leaves to a network based on the stochastic block model. We create a stochastic block model network with 2 communities, each with 500 nodes, $c_{in}=4.8, c_{out}=1.2$. We add one leaf to each node whose number of connections within the community exceeds its connections outside its community by at least 3. This selects the nodes whose degree is greater than the median and whose membership is slightly ambiguous.

\begin{figure}[htbp]
\includegraphics[width=\columnwidth]{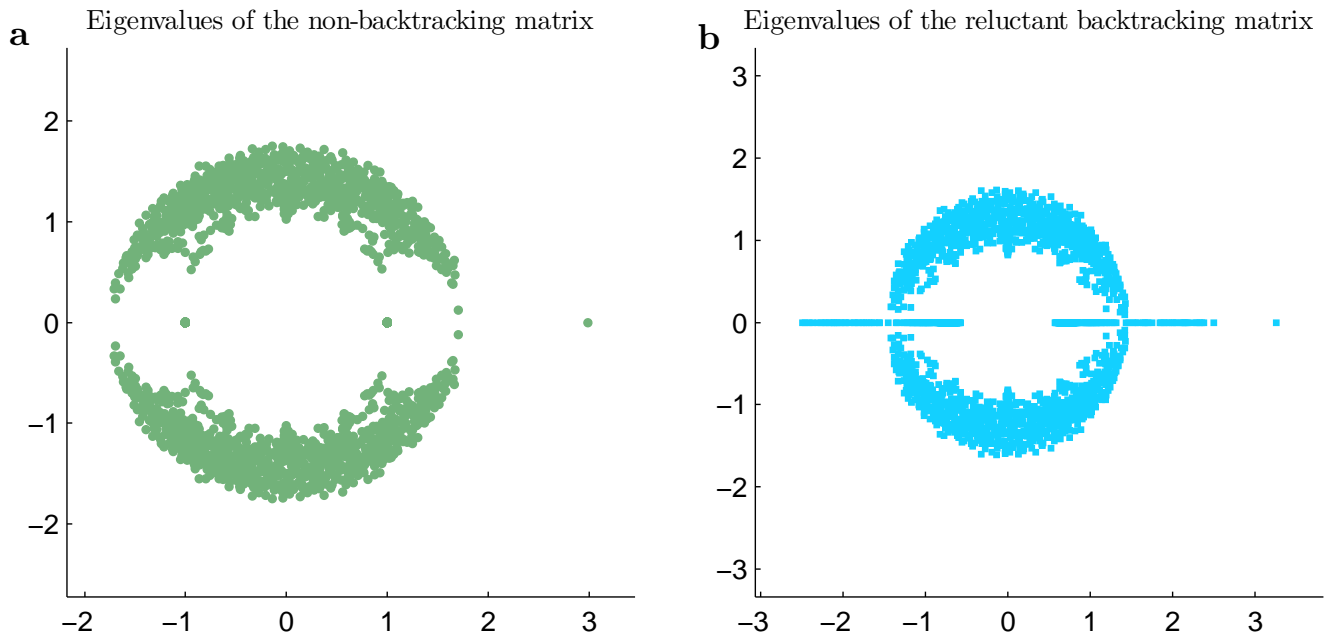}\\
\caption{Stochastic blockmodel network with additional leaves. $n=1000, m=1332, c_{in}=4.8, c_{out}=1.2$. $n$ denotes the number of nodes in the network. $m$ denotes the number of undirected edges in the network. All the random walk operators are square matrices of order $2m$. a) Eigenvalues of a representative non-backtracking matrix. Note that there is only 1 real eigenvalue outside the bulk. b) Eigenvalues of a representative reluctant backtracking matrix.}
 \label{fig:leaves}
\end{figure}

Figure~\ref{fig:leaves} shows the eigenspectrum and performance  of the non-backtracking operator $\mat{B}$ and reluctant backtracking operator $\mat{R}$ for such a network. The non-backtracking operator $\mat{B}$ does not detect two communities as there is only one real eigenvalue outside the bulk. The additional information provided by the leaves is not available to the non-backtracking operator. On the other hand, the reluctant backtracking operator accounts for the leaves in the network and its second eigenvector successfully detects two communities.

\subsection*{Stochastic block model based networks}
The quality of community detection is inversely proportional to the degree of mixing between different communities in a network. Theoretical considerations predict that performance of any spectral method falls to zero below a predictable mixing threshold for simulated networks based on the stochastic block model ~\cite{decelle2011asymptotic,decelle2011inference,mossel2012stochastic}. The networks become spectrally indistinguishable from Erd\"{o}s-R\'{e}nyi random graphs below the predicted mixing threshold, and therefore no communities can no longer be detected by spectral methods. The minimum network mixing variable, $c_{minus}$ where any spectral method can detect communities in networks based on the stochastic block model is labeled the threshold limit. Consequently, simulated networks based on the stochastic block model serve as a useful benchmark for testing the performance of different community detection methods. Krzakala et al. \cite{krzakala2013spectral} showed that the non-backtracking operator $\mat{B}$ can detect communities in sparse networks right down to the theoretical limit where other spectral methods fail.

\begin{figure}[ht] 
\includegraphics[width=0.6\columnwidth]{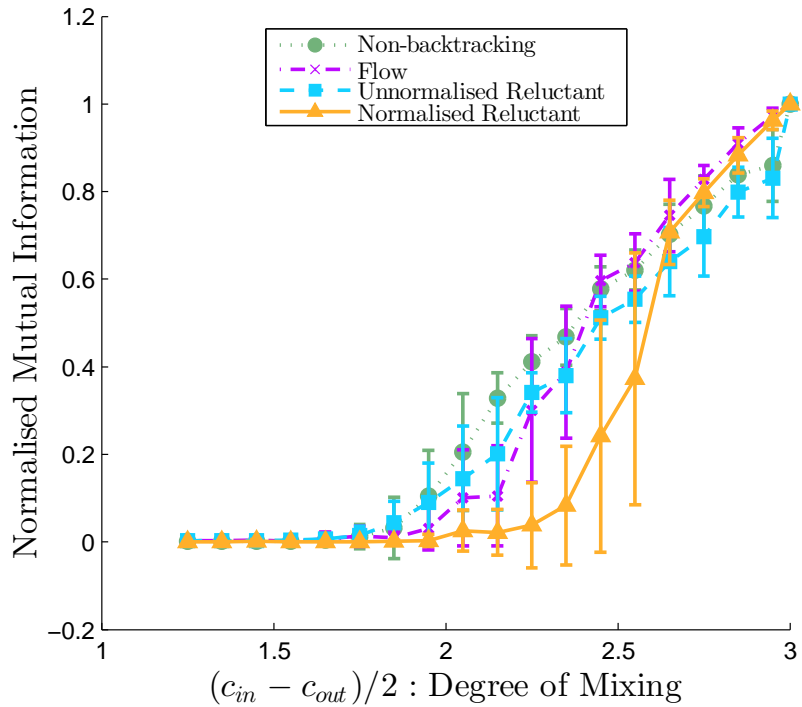}\\
\caption{Community detection performance on the stochastic block model. We plot normalised mutual information of the recovered modules compared to the planted modules as a function of the degree of mixing in the block model network (1000 nodes, average degree $c$=3). Each data point shows the mean and standard deviation of NMI for the different operators as applied to $20$ networks with the given mixing parameters. }
\label{fig:blockmodel}
\end{figure}

Figure~\ref{fig:blockmodel} shows the performance of  the four operators $\mat{B}, \mat{F}, \mat{R}$ and $\mat{P}$ on a set of networks based on the stochastic block model with $10^3$ nodes with constant average node degree and varying degrees of mixing between communities($0.1 \leq c_{minus} \leq 5.9$). The reluctant backtracking operator $\mat{R}$'s performance is comparable to the non-backtracking matrix $\mat{B}$, but the operator $\mat{P}$ performance falls to chance above the threshold limit. Thus the reluctant backtracker $\mat{R}$ accounts for hanging trees in a network, yet there is no penalty for detecting communities down to the theoretical limit.

\subsection*{Real world networks}
\begin{figure}[htbp] 
\includegraphics[width=0.85\columnwidth]{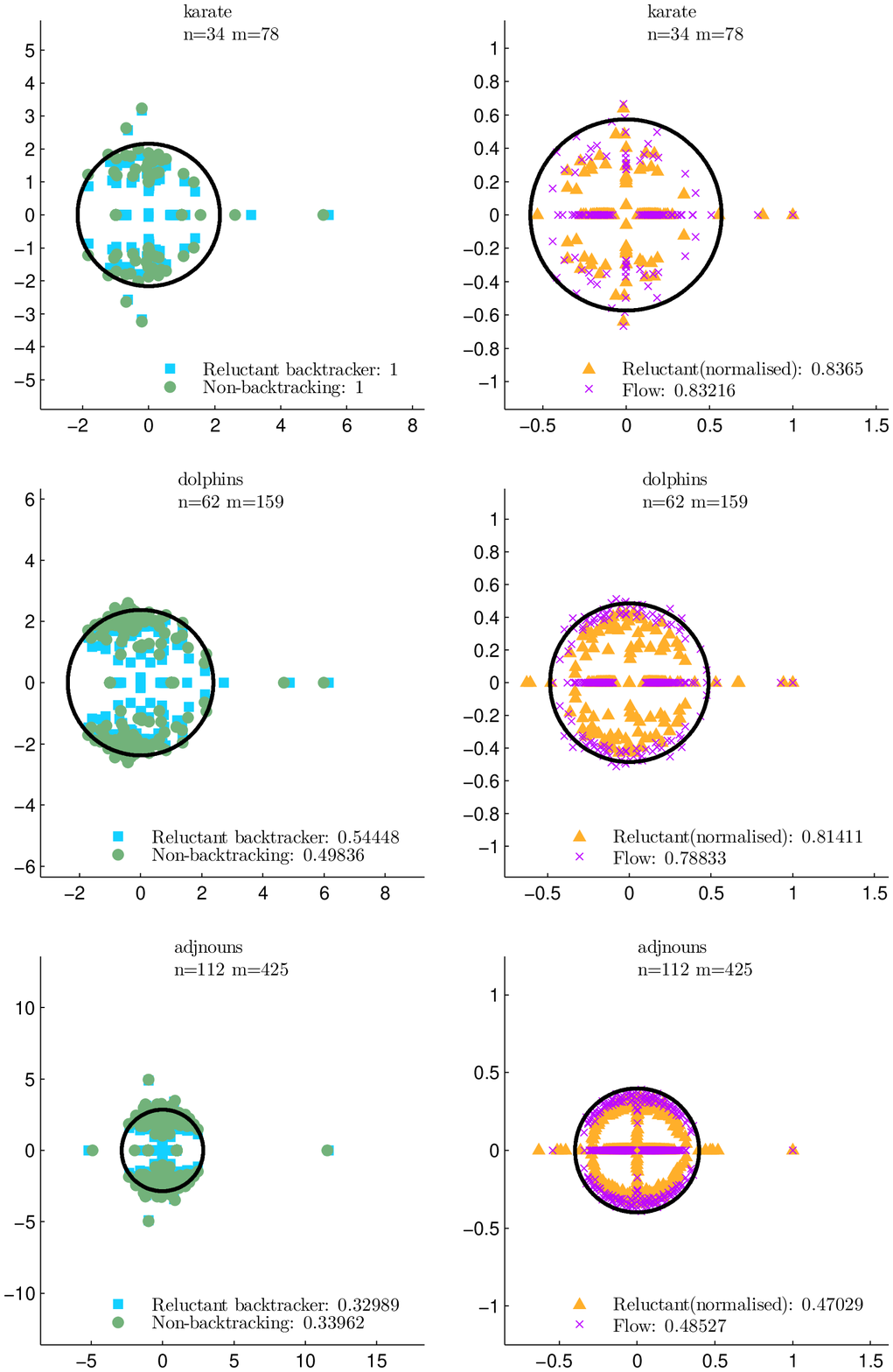}
\caption{Real world performance. The dots are the eigenvalues of the respective matrices. The black circle is the approximate analytical bound of the `bulk' eigenvalues for the non-backtracking and flow matrices, respectively $\sqrt{\langle c \rangle}$ \cite{krzakala2013spectral} and $\sqrt{\langle c / (c-1) \rangle / \langle c \rangle}$ \cite{newman2013spectral}, where $c$ is degree, and $\langle \dot~ \rangle$ is an average. These bounds were derived for the stochastic block model, so are used here as an heuristic guide for the distribution of eigenvalues resulting from the real-world networks, and computed using their degree distribution. $n$ denotes the number of nodes in the network. $m$ denotes the number of undirected edges in the network. All the random walk operators are square matrices of order $2m$. }
   \label{fig:real_data}
\end{figure}

Table~\ref{table:real_data} and Figure~\ref{fig:real_data} compares the effectiveness of the reluctant and non- backtracking matrices  on three real world data sets: Zachary karate club~\cite{zachary1977information}, dolphins~\cite{lusseau2003bottlenose} and word adjacencies~\cite{newman2006finding}. In Figure~\ref{fig:real_data} we plot the distribution of eigenvalues of each operator, showing that both the non-backtracking ($\mat{B}$, $\mat{F}$) and reluctant-backtracking ($\mat{R}$, $\mat{P}$) operators have more than one outlying eigenvalue and can thus detect community structure in these networks. The reluctant backtrackers detect communities comparably to their respective non-backtracking counterparts, and there is no loss of performance when using the reluctant matrices rather than the non-backtracking matrices. Rather, we found that the main difference in performance depended on whether or not the operators are normalised. This is particularly striking for the dolphin social network, for which the normalised operators perform similarly and both markedly better than the unnormalised versions.

\subsection*{Modularity maximisation}
Newman~\cite{newman2013spectral} showed that the second leading eigenvector of the flow matrix $\mat{F}$ maximises the widely-used modularity function $Q$ \cite{newman2006finding}, connecting the non-backtracking method to the idea of community detection as an optimisation problem. We show that the reluctant backtracking operator $\mat{P}$ also approximately optimises the modularity function $Q$.

Assume an unweighted undirected network of size $n$ with $m$ edges specified by the adjacency matrix A. The modularity function $Q$ is defined as

\begin{align}
Q &= \frac{1}{2m} \sum_{ij} \biggl[ A_{ij} - {d_i d_j\over2m} \biggr]
    \delta_{g_ig_j}\\
	A_{ij}&: \textrm{presence/absence of edge between nodes $i$ and $j$} \nonumber  \\
	d_i&: \textrm{degree of node } i \nonumber \\
	g_i&: \textrm{group membership of node } i \nonumber\\
		m&: \textrm{number of edges in the network} \nonumber
\label{eq:modularity1}
\end{align}

Following Newman's setting and notation~\cite{newman2013spectral}, assume that the network is divided into two communities and define the $n$ dimensional group membership vector $\vec{s}$ with elements $s_i \in \{-1,1\}$  denoting the membership of each node in the network. We define the quadratic form

\begin{align}
T &= \vec{u}^T(\mat{P}-\vec{1}\vec{1}^T)\vec{v}\\
 \vec{u},\vec{v}&: 2m~ \textrm{dimensional unit vectors }\vec{1} =(1,1,1,\ldots)/\sqrt{2m} \nonumber
\end{align}

If we make the particular choice $u_{i\to j} =
v_{i\to k} = s_i$, meaning that the elements of both vectors $\vec{v}$ and $\vec{u}$ are equal to
the group index of the node from which the corresponding edge {\em emerges}, then
\begin{align}
\vec{u}^T\mat{P}\vec{v} &= \sum_{\substack{\textrm{edges\ } j\to i\\
  \textrm{edges\ } l\to k}}   \bigg[ \frac{1}{d_j} \delta_{il} \delta_{jk} +
  \delta_{il}(1-\delta_{jk})  \bigg] \frac{1}{d_i -1 + \frac{1}{d_j}} s_j s_i
  \nonumber\\
 &=\sum_js_j \sum_{ik} \bigg[\frac{1}{d_i -1 + \frac{1}{d_j}} \frac{1}{d_j} A_{ik} A_{ij} \delta_{jk}+ \frac{1}{d_i -1 + \frac{1}{d_j}} (1-\delta_{jk})A_{ik} A_{ij} \bigg] s_i \nonumber\\
  &=\sum_js_j \sum_{i} A_{ij}s_i\bigg[\frac{1}{d_i -1 + \frac{1}{d_j}}( \frac{1}{d_j}  +  d_i -1) \bigg]\nonumber\\
 &=\sum_js_j \sum_{i} A_{ij}s_i \nonumber \\
 &=\vec{s}^T\mat{A}\vec{s}
\label{eq:deriv1}
\end{align}

 Also it follows that
\begin{align}
\vec{u}^T\vec{1}\vec{1}^T\vec{v} &=
   {1\over2m} \sum_{\substack{\textrm{edges\ } j\to i\\
   \textrm{edges\ } l\to k}} s_j s_i
   = {1\over2m} \sum_{ijkl} A_{ij} A_{kl} s_j s_i\nonumber\\
  &= {1\over2m} \sum_{ji} d_j d_i s_j s_i
   = \vec{s}^T {\vec{d}\vec{d}^T\over2m} \vec{s},
\label{eq:deriv2}
\end{align}

Therefore
\begin{align}
Q &= \frac{1}{2m} \vec{u}^T(\mat{P}-\vec{1}\vec{1}^T)\vec{v},\nonumber\\
&= \frac{1}{2m}\vec{s}^T(\mat{A}-\frac{\vec{d}\vec{d}^T}{2m})\vec{s}
\label{eq:modularity_eq}
\end{align}

Since the normalised reluctant backtracker $\mat{P}$ also optimises the modularity function, our spectral solution coincides with Newman's. We summarise Newman's solution here, refer to ~\cite{newman2013spectral} for further details. Solving equation~\ref{eq:modularity_eq} exactly is hard but an approximate solution can be found by standard relaxation techniques. Allow $\vec{u}$ and $\vec{v}$ to independently take any real value rather than only  $\pm 1$ and apply the constraint that $\vec{u}^T\vec{v}=2m$. This modified problem can be solved by the method of Lagrange multipliers. We get the following equation by introducing the multiplier $\lambda$ and differentiating with respect to elements of $\vec{u}$
\begin{align}
(\mat{P}- \vec{1}\vec{1}^T)\vec{v}= \lambda \vec{v}
\end{align}
The leading eigenvector of $\mat{P}- \vec{1}\vec{1}^T$ or the second leading real eigenvector of $\mat{P}$ exactly optimises the relaxed problem. We arrive at the approximate solution of the original unrelaxed problem by setting $s_i= sgn(\sum_{j} v_{i\to j})$, i.e. we sum up all the elements of the eigenvector that emerge from node $i$ and assign $s_i=1$ if the sum is positive or $-1$ if it is negative. This is very similar to the algorithm used by Krzakala et al.~\cite{krzakala2013spectral} with the difference that we sum up edges emerging from a node rather the ones incident upon it.

\section*{Discussion}
We propose a new reluctant backtracking operator to detect communities in sparse networks that accounts for hanging trees. Unlike other recent operators such as the non-backtracking matrix and the flow matrix, the reluctant backtracking operator accounts for the presence of hanging trees in a network and its eigenspectrum is shaped by their presence. We demonstrate the utility of the reluctant backtracking operator by detecting communities in simulated networks where the non-backtracking matrix is unable to do so and also show a comparable ability to detect communities in benchmark simulated and real networks.

Newman~\cite{newman2013spectral} showed that the second leading eigenvector of the flow matrix approximately maximises the modularity function by ensuring conservation of probability at each node similar to the conservation of electric current at each node in an electrical network. Following a similar argument we also show that the eigenvector of the normalised reluctant backtracking matrix $\mat{P}$ approximately maximises the modularity function.

An interesting future problem is to extend this method to reliably detect more than two communities. Determining the number of communities in a network is a problem by itself and knowing the number of communities in a network can improve the performance of community detection methods~\cite{darst2014improving}. Krzakala et al.~\cite{krzakala2013spectral} suggested a heuristic to determine the number of communities in a given network when using the non-backtracking matrix $\mat{B}$. They derived an approximate analytical bound for the uninformative eigenvalues lying inside the 'bulk' for sparse stochastic block model networks and found that the number of real-valued eigenvalues lying outside the bulk's radius served as a good heuristic to estimate of the number of modules in model networks. Newman derived a similar bound for the flow matrix $\mat{F}$ \cite{newman2013spectral}. When applied to real-world networks, a further heuristic is to compute these bounds using the mean degree of the real-world network and use them as a guide to the number of modules in that network. We plot these approximated bounds for our sample of real-world networks in Figure \ref{fig:real_data}; we note that, like the flow matrix $\mat{F}$, the eigenvalue distribution for our normalised reluctant backtracker $\mat{P}$ is particularly well-behaved with respect to the approximated bounds compared to the unnormalised matrices. We leave the determination of the bound for the reluctant operators for future work, as they do not follow simply from those derived for the non-backtracking matrices.

However, because of the approximations involved, the heuristic can fail for real~\cite{krzakala2013spectral} and simulated networks~\cite{darst2014improving}, by predicting too many real-valued eigenvalues outside the bulk and thus predicting too many modules. The optimisation of modularity $Q$ by the second eigenvector of both the flow $\mat{F}$ and normalised reluctant-backtracker $\mat{P}$ matrices suggests two further solutions for finding more than two communities. The first solution is a more cautious approach that treats the total number $q$ of real eigenvalues outside the approximated bulk radius as an upper limit for the number of communities in the network \cite{Humphries2011}. We can identify these communities by first taking each of the $q-1$ eigenvectors corresponding to the $q-1$ eigenvalues (remembering that we start from the second eigenvector) and converting them into a length $n$ vector as before -- we sum over the eigenvector entries corresponding the same source node. We can then cluster in the $\mathbb{R}^{q-1}$ space defined by these node vectors, using a standard clustering algorithm such as $k$-means: we cluster for each $k \in [2,q-1]$, and compute $Q$ for each $k$, retaining the clustering that maximises $Q$. The second solution is to apply the iterative bisection algorithm from \cite{newman2006finding}. We initially divide the network into two communities using the second leading eigenvector of $\mat{F}$ or $\mat{P}$, then iteratively divide each sub-division using the same algorithm. We compute $Q$ for each sub-division (adjusted to account for the remainder of the network \cite{newman2006finding}), stopping when $Q \leq 0$.

The difference in performance between the normalised and non-normalised versions of the operators on the real-world networks hints that normalisation is incorporating more information about the network's structure than is available to the unnormalised operator. Normalisation adds information about the degree of the transition node $i$ in the path $j\to i \to k$ to each non-zero element of the matrix of the normalised operators $\mat{F}$ and $\mat{P}$. By contrast, each path from node $j \to k$ in the non-backtracking matrix $\mat{B}$ has an equal weight of 1 irrespective of the degree of the intermediate node $i$. This new information affects the eigenspectrum of the normalised operators, and thus likely leads to the observed differences in community detection performance.  Precisely how and when this additional information is beneficial for detecting communities is presently unclear, and is the subject of future work.

\section*{Methods}
\subsection*{Normalised mutual information}
Given a network with two partitions that label the community membership of each node, normalised mutual information (NMI) quantifies the overlap between these two partitions. NMI serves as a metric to quantify the absolute performance of a community detection method and compare the relative performance of different methods.

Assume a network with $N$ nodes and partitions $\mathbb{A}$ and $\mathbb{B}$. $A_i$ is the subset of nodes in the network that belong to group $i$ in partition $\mathbb{A}$ and  $B_j$ is the subset of nodes in the network that belong to group $j$ in partition $\mathbb{B}$. Let $n_A$ and $n_B$ be the number of groups in the partitions $\mathbb{A}$ and $\mathbb{B}$ respectively. The confusion matrix $\mathbf{F}$ captures the overlap between the two partitions, its element $F_{ij}$ counts the number of nodes common to the groups $A_i$ and $B_j$. Normalised mutual information~\cite{danon2005comparing} is defined as
\begin{align}
NMI(A,B)=\frac{-2 \sum_{i=1}^{n_A} \sum_{j=1}^{n_B}F_{ij}ln(F_{ij}N/N_iN_j)}{\sum_{i=1}^{n_A}N_iln(N_i/N) + \sum_{j=1}^{n_B}N_jln(N_j/N) }
\end{align}
where
\begin{align*}
n_A,n_B &: \textrm{number of groups in partition $\mathbb{A}$ and $\mathbb{B}$ }\\
N_i,N_j &:\textrm{number of nodes in groups $A_i$ and $B_j$}
\end{align*}
NMI always lies between $0$ and $1$; $NMI=1$ only if the partitions  $\mathbb{A}$ and $\mathbb{B}$ are identical and $NMI=0$ only if the partitions  $\mathbb{A}$ and $\mathbb{B}$ are completely independent of each other.\\

%
\subsection*{Community detection algorithm and numerical considerations}

Given the adjacency matrix of a network, we first generate one of the matrices $\mat{R}~\textrm{or}~\mat{P}$. Following  Krzakala et al. \cite{krzakala2013spectral}, we calculate its second largest absolute real eigenvalue and the associated eigenvector. The eigenvector has $2m$  elements corresponding to each directed edge in the network. We group the elements of the eigenvector by the group index of the source node of each edge and sum them up to create a new vector that has $n$ elements corresponding to each node in the network. We  divide the network into two communities by grouping all nodes that have the same sign; the sign of each element represents the estimate of the reluctant backtracking operators of the node's community.

If the network has less than 500 nodes, we calculated all the eigenvalues and eigenvectors using the eig function in MATLAB. If the network was larger than 500 nodes, we first calculated the largest 50 eigenvalues by magnitude and the associated eigenvectors using the eigs function in MATLAB that is suited for sparse matrices and is based on the implicitly restarted Arnoldi iteration method~\cite{lehoucq1996deflation}. We then selected the eigenvalues whose complex part was less than $0.5~ \times ~10^{-4}$ and finally chose the eigenvalue with the second highest magnitude and its associated eigenvector.

\textbf{Author Contributions}: AS and MH  designed the study. AS analysed the data and prepared figures. AS and MH wrote the manuscript.\\

\textbf{Competing financial interests}: The authors declare no competing financial interests.

\begin{table}[htbp]
\begin{tabular}{|c|c|c|c|c|}
\hline
 & Reluctant & Non backtracking  & Normalised reluctant & Flow \\
\hline
Karate & 1 & 1 & 0.8365 & 0.8322  \\
\hline
Dolphins & 0.5445 &  0.4984 &0.8141 & 0.7883 \\
\hline
Adjnouns & 0.3299 & 0.3396 & 0.4703 & 0.4853  \\
\hline
\end{tabular}
\caption{Performance(measured as normalised mutual information) of different operators as applied to real datasets.}
\label{table:real_data}
\end{table}

\end{document}